\documentclass[fleqn,twoside]{article}
\usepackage[headings]{espcrc2}
\usepackage{graphicx}
\title{Observation of Non-Exponential Orbital Electron Capture Decays of Hydrogen-Like $^{140}$Pr and $^{142}$Pm Ions}
\author{
Yu.A.~Litvinov\address[GSI]{Gesellschaft f\"ur Schwerionenforschung GSI, 64291 Darmstadt, Germany}%
\address[JLU]{Justus-Liebig Universit{\"a}t, 35392 Gie{\ss}en, Germany}%
\thanks{Corresponding author. E-mail: y.litvinov@gsi.de}, %
F.~Bosch\addressmark[GSI], %
N.~Winckler\addressmark[GSI]\addressmark[JLU], %
D.~Boutin\addressmark[JLU], %
H.G.~Essel\addressmark[GSI], %
T.~Faestermann\address[TUM]{Technische Universit\"at M\"unchen, 85748 Garching, Germany}, %
H.~Geissel\addressmark[GSI]\addressmark[JLU], %
S.~Hess\addressmark[GSI], %
P.~Kienle\addressmark[TUM]\address[SMI]{Stefan Meyer Institut f\"ur subatomare Physik, 1090 Vienna, Austria}, %
R.~Kn{\"o}bel\addressmark[GSI]\addressmark[JLU], %
C.~Kozhuharov\addressmark[GSI], %
J.~Kurcewicz\addressmark[GSI], %
L.~Maier\addressmark[TUM], %
K.~Beckert\addressmark[GSI], %
P.~Beller\thanks{25.07.2006 deceased}, %
C.~Brandau\addressmark[GSI], %
L.~Chen\addressmark[JLU], %
C.~Dimopoulou\addressmark[GSI], %
B.~Fabian\addressmark[JLU], %
A.~Fragner\addressmark[SMI], %
E.~Haettner\addressmark[JLU], %
M.~Hausmann\address[MSU]{Michigan State University, East Lansing, Mi 48824, U.S.A.}, %
S.A.~Litvinov\addressmark[GSI]\addressmark[JLU], %
M.~Mazzocco\addressmark[GSI]\address[Pad]{Dipartimento di Fisica, INFN, I35131, Padova, Italy}, %
F.~Montes\addressmark[MSU], %
A.~Musumarra\address[CAT]{INFN-Laboratori Nazionali del Sud, I95123 Catania, Italy}%
\address[CAT2]{Universit\'a di Catania, I95123 Catania, Italy}, %
C.~Nociforo\addressmark[GSI], %
F.~Nolden\addressmark[GSI], %
W.~Pla{\ss}\addressmark[JLU], %
A.~Prochazka\addressmark[GSI], %
R.~Reda\addressmark[SMI], %
R.~Reuschl\addressmark[GSI], %
C.~Scheidenberger\addressmark[GSI]\addressmark[JLU], %
M.~Steck\addressmark[GSI], %
T.~St\"ohlker\addressmark[GSI]\address[Hei]{Ruprecht-Karls Universit{\"a}t Heidelberg, 69120 Heidelberg, Germany}, %
S.~Torilov\address[SPB]{St. Petersburg State University, 198504 St. Petersburg, Russia}, %
M.~Trassinelli\addressmark[GSI], %
B.~Sun\addressmark[GSI]%
\address[PEK]{Peking University, Beijing 100871, China}, %
H.~Weick\addressmark[GSI], %
M.~Winkler\addressmark[GSI]}%
\runtitle{Observation of Non-Exponential Orbital Electron Capture Decays of Hydrogen-Like Ions}%
\runauthor{Yu.A.~Litvinov, F.~Bosch, N.~Winckler, et al.}

\begin{document}

\begin{abstract}
We report on time-modulated two-body  weak decays observed in the
orbital electron capture of hydrogen-like $^{140}$Pr$^{59+}$ and
$^{142}$Pm$^{60+}$ ions coasting in an ion storage ring. Using
non-destructive single ion, time-resolved Schottky mass
spectrometry we found that the expected exponential decay is
modulated in time with a modulation period of about 7 seconds for
both systems. Tentatively this observation is attributed to the
coherent superposition of finite mass eigenstates of the electron
neutrinos from the weak decay into a two-body final state.
\vspace{1pc}%
\end{abstract}

\maketitle

\section{Introduction}
\par%
The accelerator facility of GSI Darmstadt with the heavy ion
synchrotron SIS coupled via the projectile fragment separator FRS
to the cooler-storage ring ESR offers a unique opportunity for
decay studies of highly ionized atoms. It is possible to produce,
separate, and store for extended periods of time exotic nuclei
with a well-defined number of bound electrons \cite{HG-PRL}. Basic
nuclear properties such as masses and lifetimes are measured by
applying the mass- and time-resolved Schottky Mass Spectrometry
(SMS) \cite{{Ra-PRL},{Li-NPA05}}.
\par%
The dependance of $\beta$-lifetimes on the atomic charge state $q$
of the parent ion has an obvious impact on our understanding of
the processes ongoing in stellar nucleosynthesis \cite{Koji}.
Several successful experiments studying weak decay of
highly-charged atoms have been performed in the past, e.g., the
experimental discovery of bound-state beta decay ($\beta_b$) at
the example of fully-ionized $^{163}$Dy \cite{Ju-PRL}.
Due to $\beta_b$ decay, fully ionized $^{187}$Re nuclei decay by 9
orders of magnitude faster than neutral atoms \cite{Bo-PRL}. The
first direct measurement of the ratio of bound and continuum
$\beta$-decay of fully-ionized $^{207}$Tl$^{81+}$ was achieved by
a direct observation of the decay and growth of the number of
parent and daughter ions using SMS \cite{Oh-PRL05}. In the course
of the present study the first measurements of orbital electron
capture (EC) in hydrogen-like (H-like) and helium-like (He-like)
$^{140}$Pr ions have been performed \cite{HG-EPJ,Li-PRL}. It was
found that the EC decay rate in H-like $^{140}$Pr ions is about
50\% higher than in He-like ions. This result including the
measured EC/$\beta^+$ branching ratios can be explained by
standard weak decay theory \cite{Pa-PRC,Ki-PLB}.
\par%
The change of the mass manifests a radioactive decay and is
evidenced by a corresponding correlated change of the revolution
frequency. The area of the Schottky frequency peak is proportional
to the number of stored ions and to the square of the atomic
charge state, $q^2$. The SMS is sensitive to {\it single} stored
ions with atomic charge states $q\ge30$ \cite{Li-NPA05}. However,
due to a large variance in determination of the peak areas, it
became apparent that only by restricting to {\it three} injected
parent ions at maximum one could exclude any uncertainty in the
determination of the {\it exact} number of circulating ions. With
this constraint the time of the decay of each stored ion can be
precisely determined. On this basis single particle
decay-spectroscopy has been developed which allows for an
unambiguous and background-free identification of a certain decay
branch \cite{HG-EPJ,Li-Frontiers}. This leads, however, to a very
laborious collection of data which requires at least some thousand
measurements to get a statistically reasonable number of decays.
%
\begin{figure}[t!]
\begin{center}
\includegraphics*[width=7.0cm]{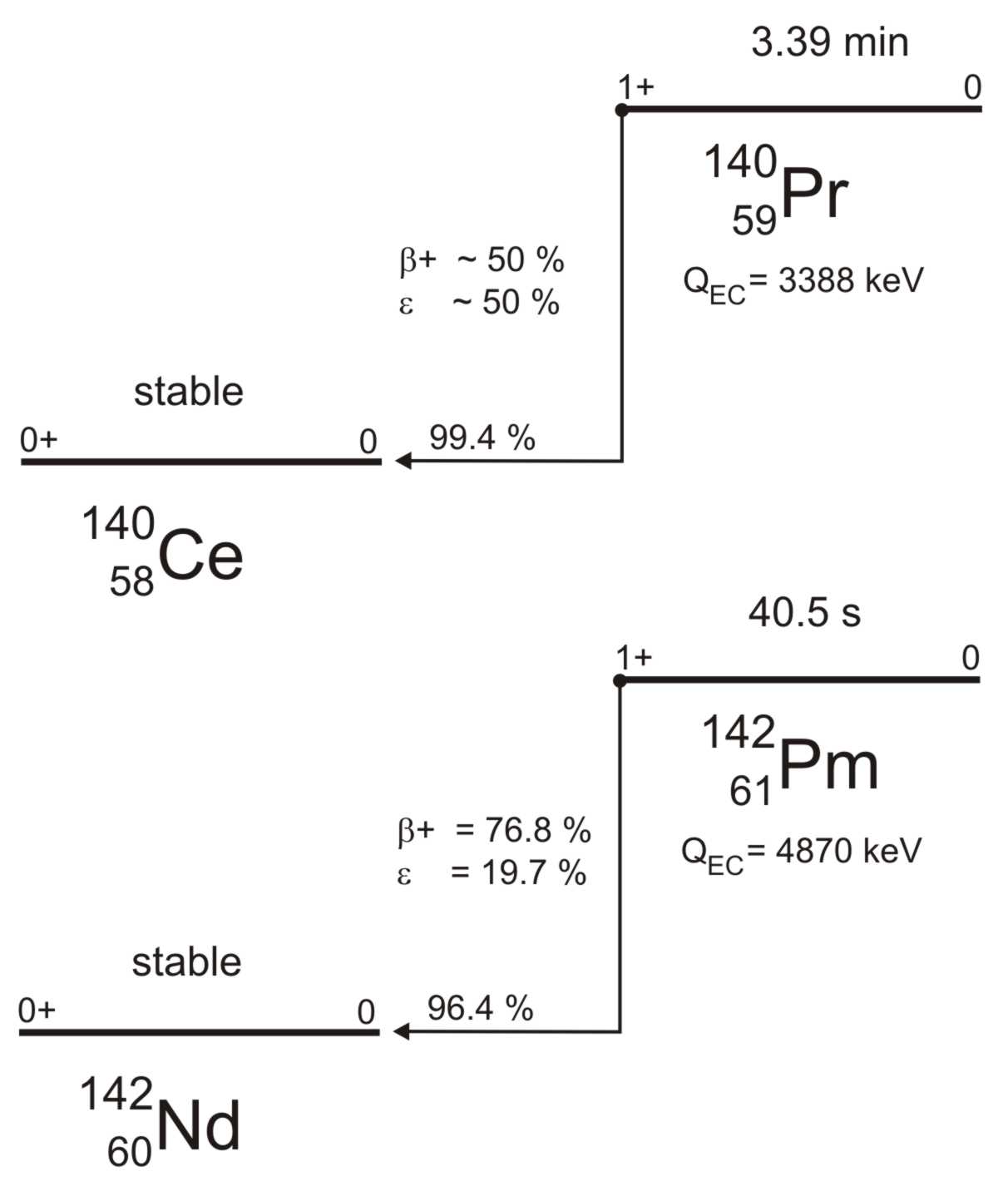}
\end{center}
\caption {Decay schemes of neutral $^{140}$Pr (upper panel) and
$^{142}$Pm (lower panel) atoms \cite{TOI}.} \label{dec-schemes}
\end{figure}
%
\par%
Here we report on the first experiments which used time-resolved
single-particle decay spectroscopy for studying the time evolution
of two-body weak decays, i.e. EC and $\beta_b$-decays of
radioactive ions in the ESR. The physics motivation was the
question whether or not the electron neutrinos generated in such
decays as coherent superposition of mass eigenstates would affect
the exponential decay \cite{LiBo-Proposal}. H-like $^{140}$Pr and
$^{142}$Pm ions have been selected for these studies. Both nuclei
decay to stable daughter nuclei via either the three-body positron
emission or the two-body EC-decay. Well-known decay schemes of
neutral $^{140}$Pr and $^{142}$Pm atoms \cite{TOI} are illustrated
in Figure \ref{dec-schemes}. Both systems decay mainly by a single
allowed Gamow-Teller ($1^+\rightarrow0^+$) transition. The weak
transitions to excited states can be safely neglected in our
context. These nuclides have quite different decay energies
($Q_{EC}$ values) and lifetimes, thus allowing a detailed
comparison of the time evolution of the decays with different
$Q_{EC}$ and lifetimes. Both $Q_{EC}$-values are sufficiently
large to be easily resolved by SMS. Furthermore, their half-lives
are much larger than the time needed for the preparation of the
ions.

\section{Experiment}
\begin{table*}[!b]
\caption{Primary beam, target and degrader parameters, number of
measurements. Each line represents a different experimental run
labelled in the first column. The ion of interest is given in the
second column. Energy of the $^{152}$Sm primary beam
$E$($^{152}$Sm) and the thickness of the beryllium production
target $L$($^9$Be) are given in the third and fourth columns,
respectively.
The number of measurements performed in each run $\#_{inj}$ is
given in the last column.}
\label{FRS_par}
\begin{center}
\begin{tabular}{|c|c|c|c|c|c|}
\hline%
 run               & ion       &$E$($^{152}$Sm)&$L$($^9$Be) &$\#_{inj}$\\
                   &           & [MeV/u]       & [mg/cm$^2$]&\\
\hline%
1&$^{140}$Pr$^{58+}$&507.8&1032&453\\
\hline%
2&$^{140}$Pr$^{58+}$&507.8&1032&842\\
\hline%
3&$^{140}$Pr$^{58+}$&601.1&2513&5807\\
\hline%
4&$^{142}$Pm$^{60+}$&607.4&2513&7011\\
\hline%
\end{tabular}
\end{center}
\end{table*}
H-like $^{140}$Pr$^{58+}$ and $^{142}$Pm$^{60+}$ ions were
produced by fragmentation of primary beams of $^{152}$Sm fast
extracted from the SIS with energies in the range between
500-600~MeV per nucleon. The duration of the extraction pulse was
less than 1~${\rm \mu{sec}}$ which is essential since we require a
well-defined time of the creation of the ions. Beryllium
production targets placed at the entrance of the FRS with
thicknesses of 1 and 2~g/cm$^2$ have been applied. The important
parameters used for the experiment are summarized in
Table~\ref{FRS_par}. Three independent experiments were performed
with Pr ions (runs 1,2 and 3) and one with Pm ions (run 4) for
comparison.
\par%
The fragments of interest were separated in-flight with the FRS
using the B$\rho$-$\Delta$E-B$\rho$ method \cite{Ge-NIM24}. For
this purpose, a 731~mg/cm$^2$ aluminum degrader was inserted at
the middle focal plane of the FRS. In runs 3 and 4 a 256~${\rm
\mu{m}}$ niobium foil after the degrader has been used in
addition. In this way we separated $^{140}$Pr$^{58+}$ and
$^{142}$Pm$^{60+}$ fragments without isobaric contaminations at
the exit of the FRS. More details on the separation of pure
$^{140}$Pr$^{58+}$ ions in these experiments can be found in Ref.
\cite{Li-Frontiers}.
\par%
Single bunches (less than 1~${\rm \mu{s}}$ long) of separated ions
containing on the average only two parent ions were injected into
the ESR at the injection energy of 400~MeV per nucleon and then
stored in the ultrahigh vacuum ($\sim10^{-11}$~mbar). Their
velocity spread caused by the production reaction was reduced
within 6-10 sec first by stochastic pre-cooling \cite{No-NIMA} and
then by electron cooling \cite{St-NIMA} to a value of $\Delta v/v
\approx 5 \cdot 10^{-7}$. The ions coasted in the ring with a
velocity of 71\% of the speed of light, corresponding to a
relativistic Lorentz factor of 1.43.
\par%
The $30^{th}$ harmonics of the revolution frequency $f$ of about
2~MHz (circumference of the ring is 108.3~m) was measured by the
Fourier frequency analysis of the signals induced by the coasting
ions at each revolution in pick-up plates. For cooled ions $f$ is
uniquely related to the mass-over-charge ratio $M/q$ of the stored
ions, which is the basis for the Schottky mass measurements
\cite{Ra-NPA}. Thus the ions of interest and their decay products
could be unambiguously identified.
\par%
The data were acquired with the commercial realtime spectrum
analyzer Sony-Tektronix 3066. It was triggered with the logic
signal corresponding to the start event of the injection kicker of
the ESR. After each trigger event, the analyzer was recording a
given number of Fourier transformed noise power spectra - FFT
(Fast Fourier Transform) frames. Each FFT frame had a bandwidth of
5~kHz and was collected for 128~msec. Each subsequent frame was
started after a defined delay of 64~msec (runs 1, 3 and 4) or
50~msec (run 2). The recorded data were automatically stored on
disk for off-line analysis.
\par%
In the ESR, the transition from the parent to the daughter ion in
a nuclear decay is evidenced by a well-defined change $\Delta{f}$
of the revolution frequency. Thus, by keeping the number of
coasting ions small we could {\it continuously} monitor the "mass"
of each ion in time and determine precisely its decay time. In our
case the parent ions could have three possible fates, namely EC or
$\beta^+$-decay or a loss due to atomic charge exchange reactions.
\par%
Since the atomic charge state $q$ does {\it not} change in the
EC-decay, $\Delta{f}$ is determined by the mass difference
($Q_{EC}$-value) between the parent and daughter nuclei. The
corresponding change in the revolution frequency is a few hundred
Hz only (about 270~Hz for the case of $^{140}$Pr and about 310~Hz
for the case of $^{142}$Pm) ($30^{th}$ harmonics). The decay is
characterized by the {\it correlated} disappearance of the parent
ion and appearance of the daughter ion. The appearance in the
frequency spectrum is delayed by about $900\pm300$~msec needed to
cool the recoiling daughter ions. Their kinetic energies are 44~eV
and 90~eV (c.m.) for the cases of $^{140}$Ce and $^{142}$Nd
daughter nuclei, respectively. The ESR lattice and the applied
ion-optical setting guarantee that {\it all} recoil ions still
remain in the acceptance irrespective on the direction of their
emission.
\par%
In $\beta^+$-decay the atomic charge changes by one unit and the
frequency of the corresponding daughter ion is shifted by about
-150~kHz ($30^{th}$ harmonics). This frequency shift is by far
larger than our small observation band, and the decay is only seen
by a decrease of the number of the parent ions. Such a
disappearance, however, cannot be distinguished from the loss of
the ion due to atomic capture or loss of an electron by reactions
with the atoms of the residual gas or the electrons of the cooler.
However, from the losses observed for the stable daughter ions a
loss constant $\lambda_{loss} \leq 2\cdot10^{-4}$~sec$^{-1}$ (the
loss constants for H-like parent ions and fully-ionized daughter
ions are almost the same \cite{Oh-PRL05}) could be extracted which
is at least one order of magnitude smaller than the EC and
$\beta^+$ decay constants $\lambda_{EC}$ and $\lambda_{\beta^+}$,
respectively. We also note that mechanical scrapers of the ESR
were positioned to remove the decay products of the atomic
charge-exchange reactions and the $\beta^+$-decay daughter ions.
%
\begin{figure*}[t!]
\begin{center}
\includegraphics*[width=10cm]{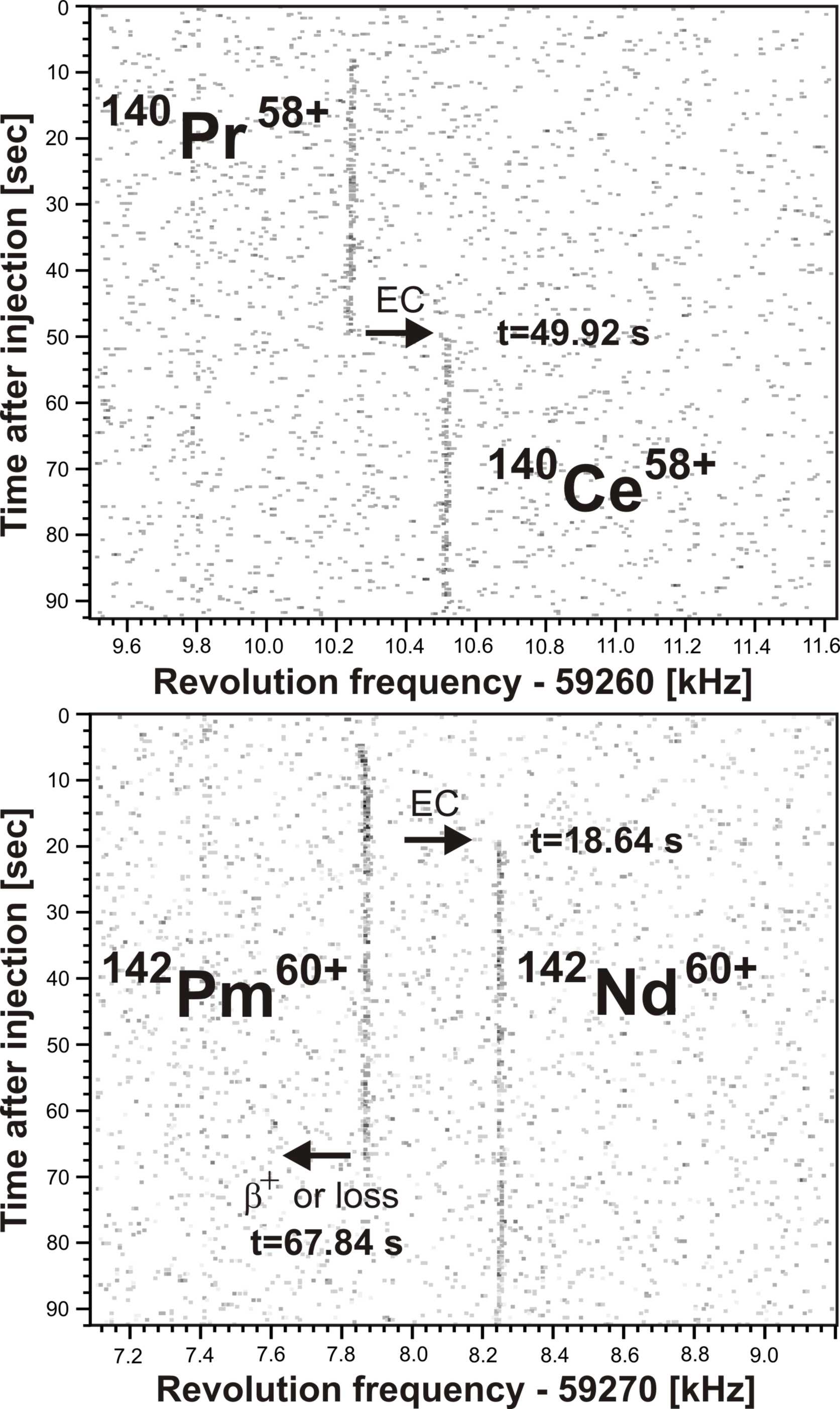}
\end{center}
\caption{\small{Upper panel: a series of consecutive frequency
spectra of a single parent $^{140}$Pr$^{58+}$ ion decaying to the
daughter $^{140}$Ce$^{58+}$ ion 49.92~sec after the injection into
the ESR. Lower panel: two injected $^{142}$Pm$^{60+}$ ions decay
18.64~sec and 67.84~sec after the injection. The first ion decays
by electron capture to a $^{142}$Nd$^{60+}$ ion. The second ion
decays by $\beta^+$-decay or is lost due to atomic charge exchange
reactions. The times of the correlated disappearance of the parent
ions and the appearances of EC-daughter ions are clearly seen. The
first few seconds are needed for cooling. The frequency
differences between parent and daughter ions correspond to
$Q_{EC}$ values of 3.35~MeV and 4.83~MeV for $^{140}$Pr$^{58+}$
and $^{142}$Pm$^{60+}$ ions, respectively.}} \label{140Pr-decay}
\end{figure*}
\par%
Two out of many thousand runs are illustrated in Figure
\ref{140Pr-decay} as a water-flow diagram starting at the time of
the injection into the ring. These examples show one (upper panel)
and two (lower panel) injected parent ions. Each horizontal line
represents a frequency spectrum with 8~Hz/channel averaged over
five consecutive FFT frames. The first several seconds are needed
for the combined stochastic and electron cooling. The decay times
are clearly seen. We emphasize that such a {\it continuous}
observation of both the parent and daughter ions {\it excludes}
any possible time-dependent alteration of the detection
efficiency.

\section{Data analysis and results}
\label{analysis}
\par%
The aim of the analysis was to study precisely the decay
characteristics of each EC-decaying ion. For this purpose, at
least two independent visual and one automatic analysis have been
applied to the data of each experimental runs.
\par%
For achieving a better signal-to-noise ratio one may average
several subsequent Schottky spectra as it is done in Figure
\ref{140Pr-decay}. In this way, however, one reduces the time
resolution. In the visual analysis we analyzed the un-averaged FFT
frames or the average over two subsequent frames. For the
automatic analysis we had to average 5 FFT frames in order to
achieve a sufficient signal-to-noise ratio. The details of the
automatic data evaluation are described in Refs.
\cite{Essel-GSI,Winckler-PHD}.
\par%
The analysis was done by inspection of each FFT spectrum taken as
a function of time. Then the time of appearance of a daughter
nucleus following the decay of its mother was determined. It was
demanded that the decay times determined in independent analysis
agree within less than one second. Only the times of the
appearance of the daughter nuclei were considered, which are
delayed compared to the decay of the mother ion by about
900$\pm$300~msec.
%
\begin{figure*}[t!]
\begin{center}
\includegraphics*[width=\textwidth]{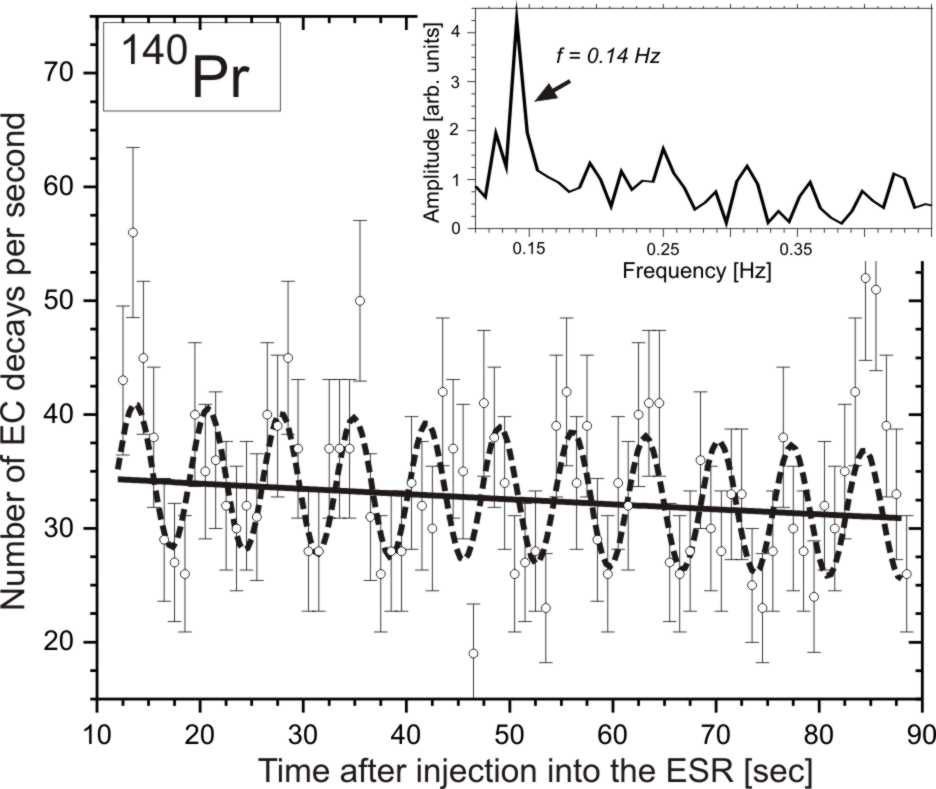}
\end{center}
\caption {Number of EC-decays of H-like $^{140}$Pr ions per second
as a function of the time after the injection into the ring. The
solid and dashed lines represent the fits according to
Eq.~\ref{exp} (without modulation) and Eq.~\ref{osc} (with
modulation), respectively. The inset shows the Fast Fourier
Transform of these data. A clear frequency signal is observed at
0.14~Hz (laboratory frame).} \label{140Pr}
\end{figure*}
%
\begin{figure*}[t!]
\begin{center}
\includegraphics*[width=\textwidth]{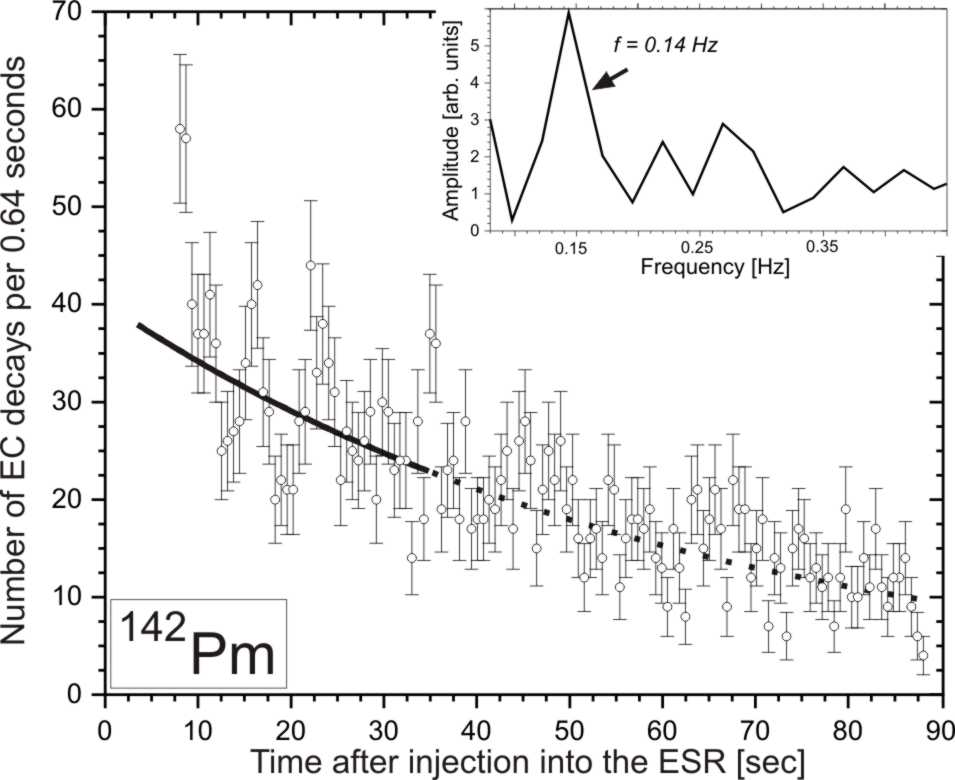}
\end{center}
\caption {Number of EC-decays of H-like $^{142}$Pm ions per 0.64
seconds as a function of the time after the injection into the
ring. The solid line represents the exponential decay fit
according to Eq. \ref{exp} until 33~sec after injection (continued
as a dotted line). The inset shows the FFT spectrum obtained from
the data until 33~sec. The reduced resolution compared to
Figure~\ref{140Pr} is explained by a smaller number of points used
for the FFT. A clear FFT peak is observed at about 0.14~Hz
(laboratory frame).} \label{142Pm}
\end{figure*}
%
\begin{figure*}[t!]
\begin{center}
\includegraphics*[width=\textwidth]{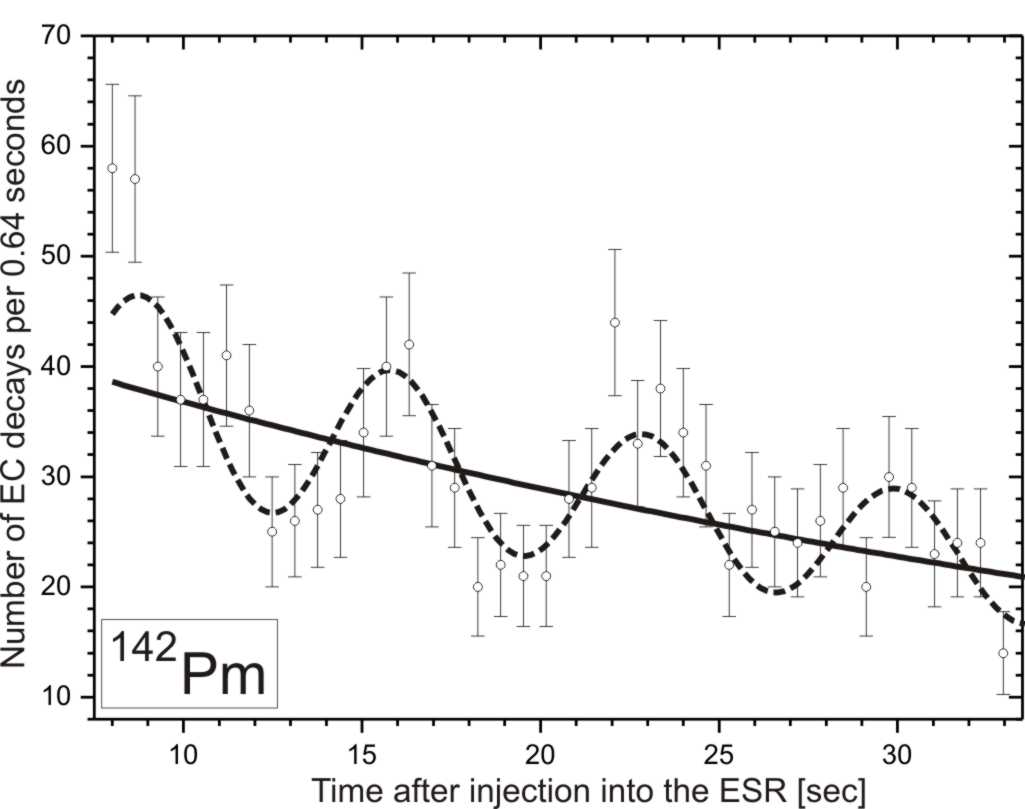}
\end{center}
\caption {A zoom to the first 33 sec after injection of the data
presented in Figure \ref{142Pm}. The solid line represents the
exponential decay fit according to Eq. \ref{exp}. The dashed line
shows the fit according to Eq.~\ref{osc}.} \label{142Pmzoom}
\end{figure*}
\par%
The decay times from the three runs with H-like $^{140}$Pr were
combined. These results and the results for $^{142}$Pm ions are
illustrated in Figure~\ref{140Pr} and in Figures~\ref{142Pm} and
\ref{142Pmzoom}, respectively. The time of the injection into the
ESR is within 1~${\rm \mu{s}}$ the time of the creation of the
ions. The data were fitted with the exponential decay function:
\begin{equation}
\frac{dN_{EC}(t)}{dt}=N(0) \cdot \lambda_{EC} \cdot e^{-\lambda
t}, \label{exp}
\end{equation}
where $N(0)$ is the number of parent ions at the time $t = 0$, the
time of injection and
$\lambda=\lambda_{EC}+\lambda_{\beta^+}+\lambda_{loss}$. The ratio
of $\lambda_{EC}/\lambda_{\beta^+}$ is 0.95(8) for the H-like
$^{140}$Pr and is expected to be about 0.32 for the H-like
$^{142}$Pm \cite{Li-PRL}.
\par%
It is clear to see that the expected exponential decrease of the
EC-decays as a function of time shows a superimposed periodic time
modulation. To account for this modulation we fitted the data with
the function:
\begin{equation}
\frac{dN_{EC}(t)}{dt}=N(0)\cdot e^{-\lambda t} \cdot
\widetilde{\lambda_{EC}}(t), \label{osc}
\end{equation}
where $\widetilde{\lambda_{EC}}(t)=\lambda_{EC} \cdot [ 1 + a
\cdot cos( \omega t + \phi )]$ with an amplitude $a$, an angular
frequency $\omega$, and a phase $\phi$ of the modulation. For the
case of $^{142}$Pm ions only the first 33 seconds after the
injection were fitted with Eq.~\ref{osc} due to the short
half-life of the mother nuclei and, thus, the fast damping of the
modulation amplitude.
\par%
The fits were done with the MINUIT package \cite{MINUIT} using the
$\chi^2$ minimization and the maximum likelyhood methods which
yielded consistent results. The fit parameters are given in
Table~\ref{ta-fits}.
\begin{table*}[!b]
\caption{The fit parameters obtained for $^{140}$Pr (upper part)
and $^{142}$Pm (lower part) EC-decay data illustrated in Figures
\ref{140Pr}, \ref{142Pm} and \ref{142Pmzoom}. The fits are done
according to Eq. \ref{exp} and Eq. \ref{osc} which is indicated in
the first column. The corresponding $\chi^2/DoF$ ($DoF$ = degrees
of freedom) are given in the last column.}
\label{ta-fits}
\begin{center}
\begin{tabular}{|c|c|c|c|c|c|}
\hline%
\multicolumn{6}{|c|}{Fit parameters of $^{140}$Pr data}\\
\hline%
Eq. & $N_0\lambda_{EC}$&$\lambda$&$a$&$\omega$&$\chi^2/DoF$\\
\hline%
\ref{exp}&34.9(18)&0.00138(10)&  -    & -     &107.2/73\\
\hline%
\ref{osc}&35.4(18)&0.00147(10)&0.18(3)&0.89(1)&67.18/70\\
\hline%
\multicolumn{6}{|c|}{Fit parameters of $^{142}$Pm data}\\
\hline%
Eq. & $N_0\lambda_{EC}$&$\lambda$&$a$&$\omega$&$\chi^2/DoF$\\
\hline%
\ref{exp}&46.8(40)&0.0240(42)&  -    & -      &63.77/38\\
\hline%
\ref{osc}&46.0(39)&0.0224(41)&0.23(4)&0.89(3) &31.82/35\\
\hline%
\end{tabular}
\end{center}
\end{table*}
\par%
From the angular frequency $\omega$ of Table~\ref{ta-fits} we can
extract the periods of the modulation of 7.06(8)~sec and
7.10(22)~sec (laboratory frame) for $^{140}$Pr and $^{142}$Pm
ions, respectively. The presence of the modulation frequencies was
also confirmed by Fast Fourier Transforms (see insets in
Figure~\ref{140Pr} and Figure~\ref{142Pm}). The amplitudes $a$
agree within the error bars. The average value of both systems is
$\langle{a}\rangle=0.20(2)$.

\section{Discussion}
\par%
The observed periodic modulations of the expected exponential
decrease of the number of EC-decays per time unit still suffer
from restricted statistics. However, the "zero hypothesis" of a
pure exponential decay can be already rejected according to the
$\chi^2/DoF$-values from Table~\ref{ta-fits} on the 99\%
confidence level (one-sided probabilities $p=0.006$) for both
investigated nuclear systems.
\par%
First of all, the finding of nearly the same oscillation period of
about 7~sec might suggest a technical artefact as their common
origin, such as periodic instabilities in the storage ring or of
the recording systems. However, this explanation is very
improbable due to our detection technique where we have--during
the whole observation time--{\it the complete and uninterrupted
information upon the status of each stored ion}. Furthermore, the
parent and daughter ions from both systems coast on different
orbits in the ESR and have different circulation times. We can
also exclude binning effects or the variance of the delay between
the decay of the mother and the "re-appearance" of the daughter
ion, since these effects lead to an uncertainty of the decay time
that is much smaller than the observed period.
\par%
It is very probable that the H-like $^{140}$Pr as well as the
$^{142}$Pm ions with nuclear spin $I=1^+$ are produced in a
coherent superposition of the two $1s$ hyperfine states with total
angular momenta $F=1/2$ and $F=3/2$. This could lead to well-known
quantum beats with a beat period $T=h/\Delta E$, where $\Delta E$
is the hyperfine splitting. However, those beat periods should be
more than twelve orders of magnitude shorter than the observed
ones.
\par%
The weak decay conserves the $F$ quantum number, and since the
final state (fully ionized daughter nuclei with $I=0^+$ and
emitted electron neutrino $\nu_e$) has $F=1/2$, the EC-decay from
the $F=3/2$ state is {\it not} allowed \cite{{HG-EPJ},{Li-PRL}}.
Only a hypothetical, yet unknown, mechanism which transfers the
parent ions periodically within 7 seconds from the $F=1/2$ ground
state to the $F=3/2$ state and back in both nuclides could
generate the observed modulations \cite{Weinheimer}.
\par%
Thus, we try to interpret the modulations as due to the properties
of the electron neutrino that is generated in the EC-decay as a
coherent superposition of at least two mass eigenstates. This
necessarily comprises that also the recoiling daughter nuclei
appear as a coherent superposition of states that are entangled
with the electron neutrino mass eigenstates by momentum- and
energy conservation.
\par%
We note that a time structure was observed in the two-body decay
of stopped pions in KARMEN experiment (KARMEN time anomaly)
\cite{KARMEN}. This anomaly was described in
Ref.~\cite{Sri-arxiv95} by using a function containing periodic
time modulation similar to Eq.~\ref{osc}.
\par%
There is a long-lasting and still persisting debate whether the
generated neutrino mass eigenstates should have the same energy or
rather the same three-momentum $\vec{q}_\nu$ or neither of them
\cite{{Bilenky},{Kayser},{Okun},{Stodolsky},{Lipkin}}. In our
case, this question can be addressed properly only in the context
of wave-packets since we observe the decaying system in a
restricted region of space and time. This necessarily generates an
uncertainty of both momentum and energy. An attempt to interpret
the modulation times in this framework has been made in
Ref.~\cite{Kienle}.
\par%
Disregarding momentum and energy spread in a simplified picture
and restricting to two neutrino mass eigenstates, one gets from
momentum and energy conservation for an initial state with energy
$E$ and momentum $P=0$ in the c.m. system \footnote{In the
following we set $c$=1.}:
\begin{eqnarray}
E_1+M+\frac{p_1^2}{2 M} = E\\
E_2+M+\frac{p_2^2}{2 M} = E,
\end{eqnarray}
where $E_i=\sqrt{p_i^2+m_i^2}$ denotes the energy of the two
neutrino mass eigenstates with masses $m_1$ and $m_2$,
respectively, $p_i^2/2M$ the corresponding kinetic energies of the
recoiling daughter nuclei, and where $M$ is the mass of the
daughter nucleus. By combining these two equations and neglecting
a term given by the ratio of the recoil energy and the mass of the
daughter nucleus we arrive at (see e.g. Ref.~\cite{Lowe,S-PLB}):
\begin{equation}
\Delta{E} = E_2 - E_1 \approx \frac{\Delta m^2}{2 M},
\end{equation}
where $\Delta m^2 = m_1^2-m_2^2$.
\par%
The modulations could be caused by the energy splitting
$\Delta{E}$ which is indicated by almost the same observed
modulation periods for both decaying nuclei $^{140}$Pr and
$^{142}$Pm with almost the same nuclear masses $M$ but with quite
different neutrino energies and, thus, momenta. One expects for a
mass of 140 mass units and for $\Delta{m}^2\approx10^{-4}$~eV$^2$
\cite{KAMLAND} a period in the c.m. system of roughly T=10~sec.
Besides the fact that this estimate is based on several
assumptions many questions remain. How could the coherence of the
entangled quantum states be preserved over time spans of some ten
seconds? What is the effect of the continuous monitoring of the
state of the ion? Is the "phase" between the entangled neutrino
mass eigenstates set back to zero at each observation?
\par%
It is obvious that our findings must be corroborated by the study
of other two-body beta decays (EC and $\beta_b$). Furthermore, it
has to be investigated how the oscillation period--if persisting
at all--depends on the nuclear mass $M$. Mandatory are also
investigations of three-body $\beta$-decays, where oscillations
should be washed out due to the broad distribution of neutrino
(sc. recoil) energies. Finally, an interesting case arises when
the decaying nucleus is not free, but couples to the full phonon
spectrum in the lattice of a solid.
%
\vspace*{+0.4cm}%
\par%
{\bf Acknowledgements}
\vspace*{+0.3cm}%
\par%
We would like to express our deep gratitude to W.~Henning for his
continuous support and invaluable advice. It is a pleasure to
acknowledge many fruitful and engaged discussions with L.~Batist,
K.~Blaum, P.~Braun-Munzinger, H.~Emling, A.~F{\"a}{\ss}ler,
B.~Franzke, S.J.~Freedman, L.~Grigorenko, A.~Ivanov, H.-J.~Kluge,
E.~Kolomeitsev, R.~Kr{\"u}cken, K.~Lindner, M.~Lindroth,
G.~M{\"u}nzenberg, Z.~Patyk, K.~Riisager, A.~Sch{\"a}fer,
J.~Schiffer, D.~Schwalm, R.~Schuch, N.~Severijns, H.~St{\"o}cker,
P.M.~Walker, J.~Wambach, and H.~Wilschut. We would like to thank
in particular H.~Feldmeier, M.~Kleber, K.H.~Langanke, H.~Lipkin,
P.~Vogel, Ch.~Weinheimer, and K.~Yazaki for intensive theoretical
discussions. We are grateful to Th.~M{\"u}ller and A.~Le F{\`e}vre
for the help in the data evaluation. We are indebted to the HADES
and IKAR collaborations for their help and flexibility concerning
the beam time schedule. The excellent support by the accelerator
team of GSI decisively contributed to the successful achievement
of our experiments. One of us (M.T.) acknowledges the support by
the A. von Humboldt Foundation.
\vspace*{-0.2cm}%

\end{document}